\documentclass{article}
\usepackage[utf8]{inputenc}
\usepackage{graphicx}
\usepackage{tabularx} % in the preamble
\usepackage{adjustbox}
\usepackage{setspace}
\usepackage{authblk}
\usepackage{hyperref}

\title{{\sc SeqMapPDB}: A Standalone Pipeline to Identify Representative 
Structures of Protein Sequences and Mapping Residue Indices in Real-Time at Proteome Scale}
\author[1]{Boshen Wang, Xue Lei, Wei Tian, Alan Perez-Rathke, and Jie Liang }
\author[2]{Yan-Yuan Tseng}
\affil[1]{Boinformatics Program, Center for Bioinformatics and Quntitative Biology, Richard and Loan Hill Department
of Biomedical Engineering, University of Illinois at Chicago, Chicago, 60612, USA. }
\affil[2]{DMolecular Medicine and Genetics, Wayne State University, Detroit, 48201, USA.}

\date{}

\begin{document}

\maketitle

\abstract{\textbf{Motivation:} 3D structures of proteins provide rich information for understanding 
their biochemical  roles. Identifying the representative protein structures 
for protein sequences is essential for analysis of proteins at proteome scale.  However, there are 
technical difficulties in identifying the representative structure of a given protein 
sequence and providing accurate mapping of residue indices. Existing databases of 
mapping between structures and sequences are usually static that are not suitable for studying proteomes with frequent gene model revisions. They often do not  provide 
reliable and consistent representative structures that maximizes sequence coverage.  
Furthermore, proteins isomers are usually not properly resolved.\\
\textbf{Results:} To overcome these difficulties, we have developed a  computational pipeline 
called {\sc SeqMapPDB} to provide high-quality representative {\sc PDB} structures of given sequences. 
It provides mapping to structures that fully cover the sequences when available, or to 
the set of partial non-overlapping structural domains that maximally cover the query sequence. 
The residue indices are accurate mapped and isomeric proteins are resolved. 
{\sc SeqMapPDB} is efficient and can rapidly carry out  
proteome-wide mapping to the selected version of reference genomes %on-the-fly 
in real-time. Furthermore, {\sc SeqMapPDB} provides the flexibility of a 
stand-alone pipeline for large scale mapping of in-house sequence and structure data.\\
\textbf{Availability:} Our method is available at \url{https://bitbucket.org/lianglabuic/seqmappdb} 
with GNU GPL license. \\
% \textbf{Contact:} \href{jliang@uic.edu}\\

\section{Introduction}
The proteomes of over 1,600 species in Eukarya and over 8,400 species in Prokarya and Archaea
have been deposited in the {\sc UniProt} database~\cite{uniprot2019}. 
Analysis of these proteomes can identify relevant
protein domains and sequence motifs to gain knowledge of the biochemical roles of the 
proteins~\cite{mistry2021pfam,linearMotif}. 
In addition, a large number of proteins have their 3D structures resolved.
The {\sc Protein Data Bank} ({\sc PDB}) currently contains the atomic coordinates of 
over 167,000 proteins, with thousands more added every year~\cite{pdb2021}.
As protein 3D structures can provide detailed information on
catalytic residues, binding pockets, and potential allostery sites important for
biochemical functions~\cite{castp, prody, namd, saltbridge}, 
incorporating structural information in analysis can reveal new insight 
into the biochemical roles of the proteins.

To incorporate protein structural information, an important task is to connect  protein sequences in the proteomes  to protein structures 
in the PDB database in an accurate and robust fashion. 
Although databases such as {\sc UniProt}~\cite{uniprot2019} and {\sc SIFTS}~\cite{sifts} do 
provide mappings between structures in the {\sc PDB} database and sequences, 
there are a number of difficulties in identifying representative 
{\sc PDB} structures and in providing accurate mapping of residue indice that 
these databases currently do not resolve.

First, mapping to a fixed sequence databases is problematic. As protein sequences are inferred from gene models, they are often 
updated once a refined gene model is validated~\cite{stanke2003gene, MSproteomics}. 
Furthermore, proteins may have several isoforms due to alternative splicing~\cite{alternative}. 
A static mapping is inadequate in these cases.
These problems are further compounded by the occasional irregular index assignment 
some protein chains have (e.g.\ chain A of {\tt 6EC3} starts at index -13)~\cite{odd_6ec3}, 
which requires correction to indice of the query sequences.

Second, many proteins have structures resolved only 
for domains that do not cover the full-length sequences, likely due to 
experimental difficulties in resolving the structures of certain protein regions~\cite{strucCoverage}.
A single protein sequence is often mapped to several non-overlapping structural domains 
and these need to be reconciled. 

Third, there are proteins whose structures have been resolved in multiple experimental studies, leading to
redundant entries in the {\sc PDB} database. It is desirable to identify the representative 
structures with maximum coverage.

Fourth, not all structures in the {\sc PDB} are of equally high quality. 
{\sc PDB} structures may often have several undetermined residues, which would cause 
mismatch to the query sequences. In addition, there are cases where 
only the coordinates of C$_\alpha$ exist,  with information of all other 
atoms missing ({\it e.g.}, PDB {\it 1W2S\/}). 
These {\sc PDB} entries with large proportion of missing atoms should be avoided.

We have developed a pipeline that can retrieve non-overlapping and high-quality {\sc PDB} structures for 
any given sequence. This pipeline works in a stand-alone fashion and can be used to provide mapping 
to the most current version of protein sequences and structures. It is efficient and can be 
employed at proteome-wide scale.

\begin{figure}[h]
    \centering
    \caption{ Overview of {\sc SeqMapPDB} Pipeline. 
              {\sc SeqMapPDB} generates a sequence database 
              of the protein structures in the {\sc PDB} using coordinate records. 
             A query sequence is then searched against this database using {\sc BLASTP}, and a set of candidate structures is then identified. 
             {\sc SeqMapPDB} then calculates the two-way identities and coverage 
             based on {\sc ClustalW2} alignment between the query sequence and each candidate sequence.  
             {\sc SeqMapPDB} then either identify the 
             full-length structure, or a set of non-overlapping partial structural 
             domains with maximal coverage, with indices mapped.}
   \includegraphics[scale=0.5]{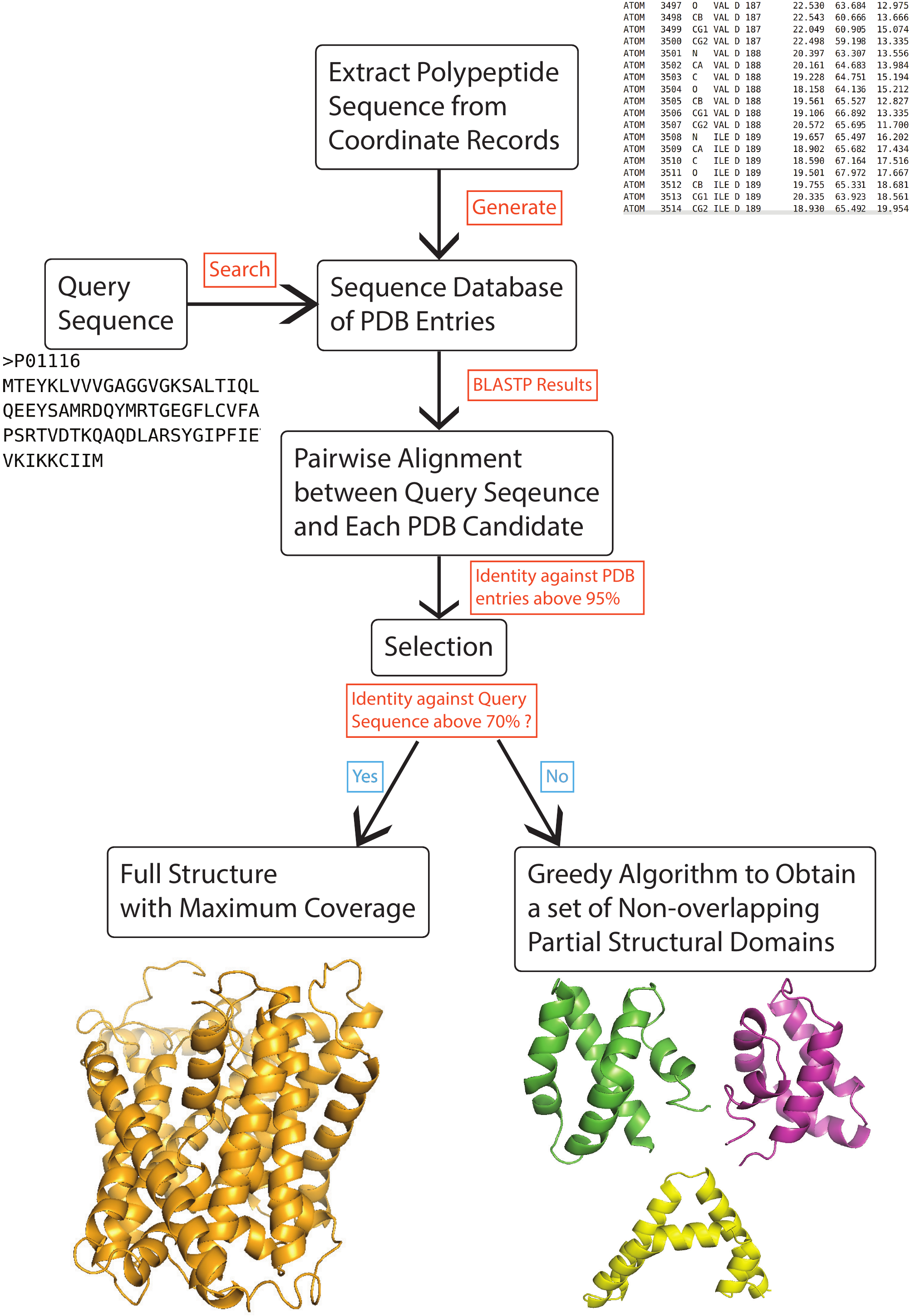}

\end{figure}

\section{Methods}

\subsection{Structural Mapping of Human Proteome}
We apply the {\sc SeqMapPDB} pipeline to the canonical forms of sequences 
from the human reference proteome (June 2021 release by {\sc UniProt}). 
Among the 20,610 proteins, 5,885 can be mapped to {\sc PDB} structures. 2,617 of 
which have full-coverage structures, and 3,268 have partially-covered structures.
763 of the partial covered proteins have structures for at least two 
non-overlapping partial domains. The full task of mapping takes about 2.2 hours 
on a machine equipped with an {\sc Intel i9-10850K} CPU, 64Gb RAM and 1Tb solid state drive.

\subsection{Mapping to Non-Overlapping Partial Structural Domains}
There are 763 proteins whose reference sequences are mapped to multiple 
non-overlapping partial domain structures.
We use the protein prosaposin ({\sc UniProt: P07602}) encoded by gene {\sc PSAP} as 
an example. Prosaposin consists of 524 amino acids, but there are no 
structures that cover its full sequence. Its sequence can be mapped to four individual 
non-overlapping structural domains. 
{\sc SeqMapPDB} maps prosaposin to the structures of
{\tt 2DOB} chain A (covering the segment of residues 60--140), 
{\tt 4V2O} chain C (residues 184--272),
{\tt 1M12} chain A (residues 311--403), and 
{\tt 3BQQ} chain A (residues 405--484). 
Overall, 64.7\% of residues in prosaposin can be mapped to structures.

Mapping provided by {\sc SeqMapPDB} produces a non-redundant set of structures 
to the sequence of prosaposin. 
In contrast, the database {\sc SIFTS} returns 33 {\sc PDB} chains for prosaposin, with many   
entries mapped redundantly to the same segment.
For example, {\sc SIFTS} produces 13 candidate {\sc PDB} structures for the segment 
of residues 405--484, 7 of which are not 100\% identical to the segment of the query sequence.
In contrast, {\sc SeqMapPDB} maps this segment to chain A of {\tt 3BQQ}, 
which is 100\% identical in sequence.

Overall, {\sc SeqMapPDB} provides mapping to non-overlapping partial structural domains of 763 proteins 
in the human proteome, where the aggregates of mapped domains cover 
from 0.7\% for long proteins (e.g. {\tt 5JDE} chain A for protein 
{\it titin\/} with 34,350 residues) to 69.9\% (e.g. {\it HLA-DRA\/}) of the full 
protein sequences, with the median coverage of 20.6\% per structural domain.

\subsection{Retrieving Paired Structures for Protein Isoforms}
A gene can encode multiple protein isoforms due to the events of alternative splicing~\cite{alternative}. 
These isoforms may fold into significantly different structures~\cite{tauisoform}.  
{\sc SeqMapPDB} can be used to identify paired structure of different isoforms.
As an illustration, the gene CRK encodes adapter molecule crk ({\sc UniProt: P46108}),
and has two isoforms deposited in {\sc UniProt}. 
Isoform 1 has 304 residues, which is mapped by {\sc SeqMapPDB} to the PDB structure of {\tt 2EYZ} (chain A),
covering the full length except one unmatched residue. 
Isoform 2 has 204 residues, which is mapped to {\tt 2EYY} (chain:A) that covers
its full length except 2 unmatched residues. 
Overall, {\sc SeqMapPDB} can obtain accurate structural mappings of 52 isoforms of 26 human proteins
with each pair of isoforms differs at least by 30 residues.

\section{Conclusion}
We have developed a standalone pipeline {\sc SeqMapPDB} that can accurately map protein sequences to 
high-quality protein structures in the {\sc PDB} database.
It can be used to select representative structures for batch processing a large 
number of query sequences, including isoforms. 
For protein sequences whose of full structures are not available, {\sc SeqMapPDB} 
can provide non-overlapping partial protein structural domains that collectively 
covers a significant portion of the query sequence. Furthermore, as protein sequences are often revised 
when reference genomes are updated, {\sc SeqMapPDB} can be employed in a stand-alone fashion
to obtain accurate and up-to-date residue index mapping.
Our pipeline is efficient, the short mapping time of the whole human proteome demonstrates 
that {\sc SeqMapPDB} can be efficiently deployed for large scale mapping of databases.
Furthermore, {\sc SeqMapPDB} provides the flexibility enabling users 
to obtain mapping between in-house databases of sequences and structures.

\section*{Funding}
This work is supported by National Institutes of Health (NIH) grants  
 R35 GM127084 and R01 CA204962.
 
\bibliography{main.bib} 

\end{document}